# The topology of the transcription regulatory network in the yeast, *S. cerevisiae*

I.J. Farkas[1,2], H. Jeong[3], T. Vicsek[2], A.-L. Barabási[3] & Z.N. Oltvai[1]

[1]Department of Pathology, Northwestern University Medical School, Chicago, IL 60611

[2]Department of Biological Physics, Eötvös University, Budapest, H-1117 Hungary

[3]Department of Physics, University of Notre Dame, Notre Dame, IN 46556

*Running head*: The organization of transcriptional activity in the yeast, *S. cerevisiae*

CORRESPONDING AUTHOR:

Zoltan N. Oltvai, M.D.
Assistant Professor
Department of Pathology
Northwestern University Medical School
Ward Bldg. 6-204, W127
303 E. Chicago Ave.
Chicago, IL 60611
Tel: (312) 503-1175
Fax: (312) 503-8240
E-mail: zno008@nwu.edu

ABSTRACT

**Motivation:** A central goal of postgenomic biology is the elucidation of the regulatory relationships among all cellular constituents that together comprise the 'genetic network' of a cell or microorganism. Experimental manipulation of gene activity coupled with the assessment of perturbed transcriptome (i. e., global mRNA expression) patterns represents one approach toward this goal, and may provide a backbone into which other measurements can be later integrated.

**Result:** We use microarray data on 287 single gene deletion *Saccharomyces cerevisiae* mutant strains to elucidate generic relationships among perturbed transcriptomes. Their comparison with a method that preferentially recognizes distinct expression subpatterns allows us to pair those transcriptomes that share localized similarities. Analyses of the resulting transcriptome similarity network identify a continuum hierarchy among the deleted genes, and in the frequency of local similarities that establishes the links among their reorganized transcriptomes. We also find a combinatorial utilization of shared expression subpatterns within individual links, with increasing quantitative similarity among those that connect transcriptome states induced by the deletion of functionally related gene products. This suggests a distinct hierarchical and combinatorial organization of the *S. cerevisiae* transcriptional activity, and may represent a pattern that is generic to the transcriptional organization of all eukaryotic organisms.

**Availability:** Detailed analyses of the comparison method and free software are available at http://angel.elte.hu/bioinf

**Contact:** vicsek@elte.hu, zno008@nwu.edu

## 1. INTRODUCTION

In the majority of single gene deletion *Saccharomyces cerevisiae* mutant strains the expression of a variable number of other genes is altered (Hughes *et al.*, 2000). This suggests the presence of a set of direct and indirect regulatory relationships among all cellular constituents that together comprise the 'genetic network' of a cell or microorganism (McAdams and Shapiro, 1995; Smolen *et al.*, 2000). The elucidation of the complete genetic network of an organism is not possible at present due to insufficient availability of microarray data and due to the fact that post-transcriptional regulatory interactions are reflected only indirectly in mRNA expression measurements. Nevertheless, experimental manipulation of gene activity coupled with the assessment of perturbed transcriptome (i. e., global mRNA expression) patterns represents an important initial approach toward this goal, and may provide a backbone into which other measurements can be later integrated (Wagner, 2001).

Here we use microarray data (Hughes *et al.*, 2000) on 287 single gene deletion *Saccharomyces cerevisiae* mutant strains (Winzeler *et al.*, 1999) to elucidate generic relationships among perturbed transcriptomes. Their comparison with a method that preferentially recognizes distinct expression subpatterns allows us to pair those transcriptomes that share localized similarities. Analyses of the resulting transcriptome similarity network identify a continuum hierarchy among the deleted genes, and in the frequency of local similarities that establishes the links among their reorganized transcriptomes. We also find a combinatorial utilization of shared expression subpatterns within individual links, with increasing quantitative similarity among those that connect transcriptome states induced by the deletion of functionally related gene products. This suggests a distinct hierarchical and combinatorial organization of the *S. cerevisiae* transcriptional activity, and may represent a pattern that is generic to the transcriptional organization of all eukaryotic organisms.

## 2. SYSTEMS AND METHODS

### 2.1. Data sets and quantitation of average transcriptome changes

Data was downloaded from Hughes *et al*, (2000) which contains two large, internally consistent, global mRNA expression subsets for the yeast, *S. cerevisiae*. One subset provides steady state mRNA expression data in wild-type *S. cerevisiae* sampled 63 separate times (the 'control' set). The other subset provides individual measurements on the genomic expression program of 287 single gene deletion mutant *S. cerevisiae* strains (Winzeler *et al.*, 1999) grown under identical cell culture conditions as wild-type yeast cells (the 'perturbation' set).

We arranged the data sets into two separate matrices as they were listed in the original data files, and containing base 10 logarithmic values. For the statistical characterization of the two matrices we use the following notations. The data matrix, ***e***, has *N* rows (each of them containing the expression levels of one gene) and *M* columns (each containing the expression levels of all genes in one microarray experiment, i.e., one measured transcriptome). The expression level of the *i*th gene in the *j*th array is $e_{ij}$, the average expression level of this gene throughout the *M* arrays is $A_i = M^{-1}\sum_{j=1}^{M} e_{ij}$ and the standard deviation of the expression level of the same gene is $\Sigma_i = \sqrt{M^{-1}\sum_{j=1}^{M}(e_{ij} - A_i)^2}$ . The average expression level of genes in the *j*th array is $a_j = N^{-1}\sum_{i=1}^{N} e_{ij}$ and the standard deviation of the expression level in the same array is $\sigma_j = \sqrt{N^{-1}\sum_{i=1}^{N}(e_{ij} - a_j)^2}$ .

### 2.2. Correlation search method

To search for correlations among transcriptomes, we compared each pair of transcriptomes individually. For any given transcriptome pair, first we identified the list of genes with known expression level values in both transcriptomes. (In the prepared data file, we called a value known, if it was not missing and was not +2 or –2, the latter indicating an experimental cutoff).

Next, we defined a segment (i.e., a small subset of the transcriptomes) with size *s*, and jump *t*, both equal to *30*, (see the Supplementary Material for analyses with other parameters). We placed a segment on the first *s* genes with known expression values in both transcriptomes. The two data sets to be compared are now the 1., 2., …, *s*. gene expression level values of the first selected transcriptome and the 1., 2., …, *s.* gene expression level values of the second selected transcriptome. We denoted these two sets (two vectors) by $\mathbf{e_1}=\{e_{1,1}, e_{2,1}, \ldots, e_{s,1}\}$ and $\mathbf{e_2}=\{e_{1,2}, e_{2,2}, \ldots, e_{s,2}\}$, respectively. Next, we computed the mean values ($M_1$ and $M_2$) and standard deviations ($\sigma_1$ and $\sigma_2$) of these two sets: $M_1 = s^{-1}\sum_{j=1}^{s} e_{j,1}$ and: $\sigma_1 = \sqrt{s^{-1}\sum_{j=1}^{s}\left(e_{j,1} - M_1\right)^2}$ ($M_2$ and $\sigma_2$ were obtained similarly).

For the measure of similarity between the two segments, $\mathbf{e_1}$ and $\mathbf{e_2}$, we used the absolute value of the (Pearson) correlation: $C_{1,2} = \left|(s\sigma_1\sigma_2)^{-1}\sum_{j=1}^{s}(e_{j,1} - M_1)(e_{j,2} - M_2)\right|$. Next, the segment of length *s* was shifted multiple times by steps of *t*, and $C_{1,2}$ was computed for the segment containing the genes *t*, *t*+1, …, *t*+*s*, then for the segment with the genes 2*t*, 2*t*+1, …,2*t*+*s*, etc. Except where explicitly mentioned, the step size is equal to the length of the segment: *t*=*s*. The similarity score between the two selected transcriptomes was defined as the $m$=10$^{th}$ largest $C_{1,2}$ value measured for them. On the resulting graph two nodes were connected, if the similarity score computed for the two transcriptomes they represent exceeded a fixed $C_0$ threshold. Note, that while the three parameters: *s*, *m* and $C_0$ are preassigned, changing the values of *s*, *m* and $C_0$ , or randomly reordering the genes' listing will not alter the essential features of the observed network. Also, after scrambling the expression values in each transcriptome independently (i.e., removing any potential correlations between the transcriptomes), the stepwise similarity search method does not identify any links, confirming that the uncovered transcriptome similarity network is not a numerical artifact of the algorithm (see the Supplementary Material for additional details).

### 2. 3. Spectral analysis

The adjacency matrix of a graph, *G*, with *N* vertices is an *N x N* symmetric matrix, ***A***, where $A_{ij}$=1 or $A_{ij}$=0, if the vertices *i* and *j* are connected, or not, respectively. Diagonal entries are 0: $A_{ii}$=0 for each *i*. The spectrum – i.e., the set of eigenvalues – of the graph's adjacency matrix, ***A***, is also called the spectrum of the graph, *G*, itself (Cvetkovic *et al.*, 1980). The inverse participation ratio of a normalized eigenvector of *G* is the sum of the fourth powers of the components of that eigenvector. Localized eigenvectors can be identified by their high inverse participation ratios. On the other hand, highly localized eigenvectors indicate the structural predominance of a handful of vertices on the graph *G*, and therefore they can be used to detect various graph structures even for small graphs with only a few hundred vertices (Farkas *et al.*, 2001). Further details are provided in the Supplementary Material.

### 2.4. Statistical characterization of the transcriptome similarity network

For the analysis displayed on Figure 3a, for each gene product the following fields of its YPD (Costanzo *et al.*, 2000) entry were used: Cellular Role, Biochemical Function, Molecular Environment and Subcellular Localization. We first analyzed each pair of the 287 transcriptomes separately using the cellular roles of the products of the two deleted genes (many possessing more than one cellular role). The union (i.e., the cellular role categories on at least one of the lists) and the intersection (i.e., the cellular role categories on both lists) of the two lists were created. We defined the identity, *I*, of the two lists, as the ratio of the number of items in the intersection vs. the union. If the union contained no categories (or only the category "unknown"), i.e., none of the two gene products had a known category, we ignored this transcriptome pair.

At each similarity threshold value, *C*, Figure 3a displays the average *I* value for those transcriptome pairs that the stepwise similarity search method predicted to be coupled stronger than *C*. This test was performed for all four databases separately.

For each adjacent pair of the similarity graph's links the 10 transcriptome segments establishing the two links were listed. The identity, *I*, of these two lists is shown on Figure 4b. Similarly, the 10 genes with the highest contributions to the two links were listed, and the identity of these lists was computed. Here, the contribution of a gene to a link denotes the absolute value of the product of the expression values of the gene in the two connected transcriptomes.

## 3. RESULTS AND DISCUSSION

To begin uncovering important generic characteristics of transcriptional organization, we assessed the degree of similarity among the genomic expression program of 287 single gene deletion mutant *S. cerevisiae* strains (Hughes *et al.*, 2000). Initial statistical analyses indicated, that compared to the wild-type yeast transcriptome, on average the expression level of only about one tenth of all genes were affected (see the Supplementary Material and http://angel.elte.hu/bioinf for details). However, most current mathematical algorithms compare transcriptomes based on their global properties thereby missing more subtle local relationships. Moreover, the analysis of singly measured transcriptomes is hampered by the observed inherent fluctuations in gene expression levels (see Hughes *et al*, (2000), and the Supplementary Material). Therefore, we introduced an analytical approach that both attenuates the effect of gene expression fluctuations and is preferentially sensitive to the recognition of local similarities among transcriptomes.

The method, illustrated in Figure 1a, divides each transcriptome into *L* short segments and systematically searches each pair of transcriptomes for similar expression patterns on all *L* segments. For transcriptomes *i* and *j*, first we sequentially determine the absolute value of the correlation coefficient, *C*, between the same *k*th segment of transcriptomes *i* and *j*. If we find at least *m* segments with correlation coefficients exceeding in absolute value a pre-assigned $C_0$ similarity parameter, we then consider transcriptome *i* and *j* to be locally similar and denote this relationship by connecting them with a link. Increasing the value of $C_0$ will increasingly limit connections to highly correlated transcriptome pairs. Decreasing $C_0$ will gradually connect more weakly similar transcriptome pairs as well, resulting in an increase in both the number of connected transcriptomes and the density of links among them (Figure 1b).

Ultimately, the totality of the links creates a similarity network in which each node represents one of the 287 deleted genes and their corresponding transcriptional response programs. For example, in Figure 1c, the detailed topology of the similarity network is shown for $C_0$ = 0.8, which corresponds to links among transcriptome states that are ≥ 80% similar in their ten most similar segments. At this similarity level we find that ~40% of the perturbed transcriptomes (113 out of 287) are linked to each other, the most highly connected transcriptomes often forming easily discernable loops within a large, central cluster (Figure 1d). In contrast, when two transcriptomes are connected only to each other, but are disconnected from all other components (Figure 1c), they share highly specific response similarities likely to be related to the specific effect of their perturbations.

To start deciphering the detailed relationships among the deleted *S. cerevisiae* genes, we first assessed the large-scale features of the similarity network's topology. We initially created three idealized test graphs to compare them with the largest cluster of the measured graph. The test graphs include an uncorrelated random (Erdos and Renyi, 1960)-, a small-world (Watts and Strogatz, 1998)-, and a scale-free graph (Barabasi and Albert, 1999), representing the three major network families known in graph theory (Strogatz, 2001). Figure 2a depicts the descending sequence of connectivities for the transcriptome graph and the three test graphs, and Figure 2b-d display the inverse participation ratios of the graph's eigenvectors vs. the corresponding eigenvalues, a measure that is known to be sensitive to the graph's topology even for small graphs (Farkas *et al.*, 2001). At all similarity levels we find that the scale-free test network's connectivity distribution (Figure 2a) and its spectral properties (Figure 2d) practically overlap with that of the *S. cerevisiae* transcriptome graph, a topology that is apparently also shared by the transcriptome similarity network of *Caenorrhabditis elegans* (Kim *et al.*, 2001) (Supplementary Material). From a biological point of view, this demonstrates that the deletion of certain gene products elicits transcriptional profiles with a significant number of expression subpatterns induced very similarly among various other perturbed transcriptomes. It also suggests a potential regulatory relationship among their corresponding genes such, that the ones possessing many shared expression subpatterns directly or indirectly regulate those that contain comparatively fewer (Wagner, 2001). Moreover, it shows that the observed similarity relationships self-organize into a continuum hierarchy in such a way, that of nodes/ transcriptomes with decreasing connectivity increasingly higher numbers occur.

To further understand the position of individual nodes/ deleted genes within the similarity network, we first examined the relationship between any two connected transcriptomes and the biochemical and cellular characteristics of their corresponding gene products, according to their categorization in the Yeast Protein Database (YPD) (Costanzo *et al.*, 2000). As shown in Figure 3a, we find that with increasing $C_0$ similarity threshold there is an increased likelihood that the connected transcriptomes represent gene products with an identical cellular role, biochemical function, molecular environment and subcellular localization. We observe, however, that to an extent local similarities are also shared among transcriptome pairs whose corresponding gene products participate in unrelated cellular activities, thus suggesting a conserved utilization of expression subpatterns.

We also determined the identity and cellular role of the corresponding gene products for the most highly connected transcriptomes. In a decreasing order of connectivity Figure 3b lists the 25 most connected nodes/ deleted gene products at various $C_0$ threshold values. Note, that the decreasing order of connectivity for the linked transcriptomes are not completely independent of $C_0$, yet many of the same nodes with only slightly modified order appear as most connected for a broad range of $C_0$ values. Specifically, the deletion of ymr031w-a, yhl029c (genes with unknown function), yel008w (stress response), gcn4 (transcriptional activator), sir2 (histone deacetylase), and swi4 (transcription factor) elicits transcriptional responses that contain the highest number of shared expression subpatterns, irrespective of the stingency of similarity. A similar trend with a lower number of shared subpatterns, is observed upon the deletion of e.g., erg2, erg3 and yer044c (ergosterol biosynthesis). In contrast, the deletion of gene products with mitochondrial functions (yer050c, msu1, rml2) elicits expression subpattern changes that are shared at a high stringency level of similarity with each other, but disproportionately less with those transcriptomes that are induced by the deletion of genes with unrelated functions. Thus, irrespective of the chosen similarity threshold, the deletion of transcriptional activators, global regulators of chromosome state, and those with a potential to induce stress response (e.g., through changes in membrane lipid composition (Bammert and Fostel, 2000)) appear to elicit the largest number of shared expression subpatterns.

Links among paired transcriptomes are established through the combinations of various transcriptome segments prompting us to assess them and their most prominent genes. To appraise the segment composition of individual links we calculated the fraction of shared segments between all pairs of links connected to the same transcriptome. We find that those pairs of links that are established at a higher stringency of similarity between any three nodes share an increasing number of identical segments (Figure 4b). Yet, it is apparent that on average the number of shared segments don't exceed more than ~ 40% of all segments. There is also a substantial statistical variability in such a way that for high-confidence loops within the large, central cluster (see e.g., Figure 1d) such similarities occur more frequently, a pattern that is highly similar for the most dominant genes within all pairs of adjacent connections (Figure 4b).

We also quantified the participation of individual segments within all links, and observed that their distribution follows a power-law with an exponent close to $\gamma = 3$ (Figure 4a). This indicates that in their totality shared expression subpatterns participate in establishing links along a continuum hierarchy from a few of them participating in many connections (the most stereotypic similarities) to many being shared among only a few transcriptomes (the most specific similarities). To identify and characterize the most prominent genes within all similarity links we first selected the ten most common segments that participate in connecting the various nodes (Figure 4c, left column). Next, for each of these segments, we determined the ten genes with the strongest overall contribution to the coupling of all linked transcriptome pairs. As shown in Figure 4c, there is a significant variability in the cellular role of genes among the different segments, the highest percentage being those with unknown function (10-60% in all 10 segments). In general, however, there are many that plays a role in stress response, various aspects of RNA- and protein metabolism or in other metabolic processes, a pattern that is similar to that observed in yeast cells upon various environmental challenges (Causton *et al.*, 2001; Gasch *et al.*, 2000).

The elucidation of the complete genetic network of *S. cerevisiae* is not possible at present due to available microarray data being restricted to a limited number of single gene deleted strains (Hughes *et al.*, 2000), the continued refinement of its genome (Kumar *et al.*, 2002), and by the fact that post-transcriptional regulatory interactions are reflected only indirectly in mRNA expression data (Wagner, 2001). Yet, our comparison and analyses of the expression subpatterns of 287 various perturbed *S. cerevisiae* transcriptomes enabled us to uncover important insights into the framework of its organization on a transcriptional level. Notably, with a novel, cut-off based method we identify a continuum hierarchy in the regulatory relationship among the yeast transcriptional elements that as a whole suggests a robust and error-tolerant scale-free topology (Barabasi and Albert, 1999) of the *S. cerevisiae* genetic network. There is the additional finding of a distinct combinatorial utilization of expression subpatterns, which in their totality also display a continuum hierarchy in their participation frequency and whose shared similarities are proportional to the functional relatedness of their corresponding gene products.

In agreement with our result, Featherstone and Brodie (Featherstone and Broadie, 2002) have recently demonstrated that besides the well-known statistical and comparative methods, random graph theory is also a powerful tool for the analysis of large scale gene perturbation experiments. They used a simple statistical method (built on P values) to create a directed network of the genes in the same data set that we have used in the present study. The underlying undirected graph was found to display a power-law behavior in the connectivity of nodes, which is a fingerprint of scale-free networks. A mathematically more sound statement, but only a prediction, concerning the same data set has been made by Wagner (Wagner, 2002). In this work a directed network of genes was hypothesized by a more careful analysis of the statistical properties of the data. The unnecessary elimination of the "noisy" values from the data set by P tests, as above, would have meant the removal of important information buried under noise. Moreover, in Wagner's work the directionality of the network has been properly taken into account. Also, it is pointed out that a complete power-law behavior cannot hold for any distribution derived from real data, only if a cutoff is included in the description.

Biological activities within *S. cerevisae* are thought to arise from shared utilization of its proteome comprised mostly of protein complexes with a conserved core and transient edges (Gavin *et al.*, 2002; Ho *et al.*, 2002). Together with other regulatory interactions, transcriptional activities play a pivotal role in establishing these dynamic compositions according to developmental states and environmental effects. The combination of microarray data with the presence of known and putative regulatory motifs in the promoter regions of the expressed genes (Bussemaker *et al.*, 2001; Pilpel *et al.*, 2001) suggests the combinatorial activity of a small number of transcription factors are responsible for a complex set of expression patterns under diverse conditions (Pilpel *et al.*, 2001). Our demonstration of a continuum hierarchy of transcriptional regulatory relationships with a seemingly conserved but malleable transcriptional output is compatible with this type of regulation.

**Acknowledgements**

We would like to acknowledge S. Friend and colleagues at Rosetta Inpharmatics (Hughes *et al.*, 2000) for making their database publicly available for the scientific community. We also thank X. He for performing the $\chi^2$ test of gene expression distributions. Research at Eötvös University was supported by the Hungarian National Research Grant Foundation (OTKA), and by the Department of Energy and the National Institute of Health at the University of Notre Dame and at Northwestern University.

## References

Bammert, G. F., and Fostel, J. M. (2000). Genome-wide expression patterns in Saccharomyces cerevisiae: comparison of drug treatments and genetic alterations affecting biosynthesis of ergosterol. *Antimicrob Agents Chemother* **44**, 1255-1265.

Barabasi, A.-L., and Albert, R. (1999). Emergence of scaling in random networks. *Science* **286**, 509-512.

Bussemaker, H. J., Li, H., and Siggia, E. D. (2001). Regulatory element detection using correlation with expression. *Nat Genet* **27**, 167-171.

Causton, H. C., Ren, B., Koh, S. S., Harbison, C. T., Kanin, E., Jennings, E. G., Lee, T. I., True, H. L., Lander, E. S., and Young, R. A. (2001). Remodeling of yeast genome expression in response to environmental changes. *Mol Biol Cell* **12**, 323-337.

Costanzo, M. C., Hogan, J. D., Cusick, M. E., Davis, B. P., Fancher, A. M., Hodges, P. E., Kondu, P., Lengieza, C., Lew-Smith, J. E., Lingner, C.*, et al.* (2000). The yeast proteome database (YPD) and Caenorhabditis elegans proteome database (WormPD): comprehensive resources for the organization and comparison of model organism protein information. *Nucleic Acids Res* **28**, 73-76.

Cvetkovic, D. M., Doob, M., and Sachs, H. (1980). Spectra of Graphs (New York).

Erdos, P., and Renyi, A. (1960). On the evolution of random graphs. *Publ Math Inst Hung Acad Sci* **5**, 17-61.

Farkas, I. J., Derenyi, I., Barabasi, A.-L., and Vicsek, T. (2001). Spectra of "real-world" graphs: beyond the semi-circle law. *Physical Review E* **64**, 026704:026701-026712.

Featherstone, D. E., and Broadie, K. (2002). Wrestling with pleiotropy: genomic and topological analysis of the yeast gene expression network. *Bioessays* **24**, 267-274.

Gasch, A. P., Spellman, P. T., Kao, C. M., Carmel-Harel, O., Eisen, M. B., Storz, G., Botstein, D., and Brown, P. O. (2000). Genomic expression programs in the response of yeast cells to environmental changes. *Mol Biol Cell* **11**, 4241-4257.

Gavin, A. C., Bosche, M., Krause, R., Grandi, P., Marzioch, M., Bauer, A., Schultz, J., Rick, J. M., Michon, A. M., Cruciat, C. M.*, et al.* (2002). Functional organization of the yeast proteome by systematic analysis of protein complexes. *Nature* **415**, 141-147.

Ho, Y., Gruhler, A., Heilbut, A., Bader, G. D., Moore, L., Adams, S. L., Millar, A., Taylor, P., Bennett, K., Boutilier, K.*, et al.* (2002). Systematic identification of protein complexes in Saccharomyces cerevisiae by mass spectrometry. *Nature* **415**, 180-183.

Hughes, T. R., Marton, M. J., Jones, A. R., Roberts, C. J., Stoughton, R., Armour, C. D., Bennett, H. A., Coffey, E., Dai, H., He, Y. D.*, et al.* (2000). Functional discovery via a compendium of expression profiles. *Cell* **102**, 109-126.

Kim, S. K., Lund, J., Kiraly, M., Duke, K., Jiang, M., Stuart, J. M., Eizinger, A., Wylie, B. N., and Davidson, G. S. (2001). A gene expression map for Caenorhabditis elegans. *Science* **293**, 2087-2092.

Kumar, A., Harrison, P. M., Cheung, K. H., Lan, N., Echols, N., Bertone, P., Miller, P., Gerstein, M. B., and Snyder, M. (2002). An integrated approach for finding overlooked genes in yeast. *Nat Biotechnol* **20**, 58-63.

McAdams, H. H., and Shapiro, L. (1995). Circuit simulation of genetic networks. *Science* **269**, 650-656.

Pilpel, Y., Sudarsanam, P., and Church, G. M. (2001). Identifying regulatory networks by combinatorial analysis of promoter elements. *Nat Genet* **10**, 10.

Smolen, P., Baxter, D. A., and Byrne, J. H. (2000). Mathematical modeling of gene networks. *Neuron* **26**, 567-580.

Strogatz, S. H. (2001). Exploring complex networks. *Nature* **410**, 268-276.

Wagner, A. (2001). How to reconstruct a large genetic network from n gene perturbations in fewer than n(2) easy steps. *Bioinformatics* **17**, 1183-1197.

Wagner, A. (2002). Estimating coarse gene network structure from large-scale gene perturbation data. *Genome Res* **12**, 309-315.
Watts, D. J., and Strogatz, S. H. (1998). Collective dynamics of 'small-world' networks. *Nature* **393**, 440-442.
Winzeler, E. A., Shoemaker, D. D., Astromoff, A., Liang, H., Anderson, K., Andre, B., Bangham, R., Benito, R., Boeke, J. D., Bussey, H., *et al.* (1999). Functional characterization of the S. cerevisiae genome by gene deletion and parallel analysis. *Science* **285**, 901-906.

**Figure legends**

**Figure 1** The transcriptome similarity search method.

***a***, Schematic display of a hypothetical microarray data set with 3 experiments (*e1*-*e3*), and 50 genes. On the five gene segments of 10 genes each, the three experiments are similar to a different extent, as indicated on the right. In the first segment there is a high similarity between all three experiments. The second segment displays similarity only between *e1* and *e2*, while the expression values of the genes in the fourth segment are highly dissimilar. ***b***, Color plot of the transcriptome similarity network at the indicated $C_0$ similarity thresholds. Each node represents a transcriptome and two transcriptomes are connected if they contain sufficient numbers of local similarities in their genomic expression patterns. Links between nodes are colored according to the similarity level between the two connected transcriptomes; green ($0.8 < C < 0.84$), yellow ($0.84 < C < 0.88$), orange ($0.88 < C < 0.92$) and red ($C > 0.92$) are used. ***c***, Enlarged view of the graph obtained for $C = 0.8$. Each node is labeled with the name of the deleted gene/experiment (Hughes *et al.*, 2000). ***d***, A detailed diagram showing four highly connected nodes (marked with white in Figure 1*c*) and five high-confidence links ($C > 0.92$) among them, with the ten most dominant genes coupling a pair of experiments listed for each connection. Those involved in all five connections are shown in red.

**Figure 2** The topological characterization of the similarity graph's central component.

***a***, Connectivity distribution for linked transcriptomes (black) vs. an uncorrelated random- (blue), a small-world- (green), and a scale-free graph (red) at $C= 0.7$. In the test graphs, the number of links and nodes are the same as in the measured graph. ***b-d***, Spectral comparison of the measured graph and the three test graphs.

**Figure 3** The comparison of the deleted genes with connected transcriptome states.

***a***, The average identity of the category lists of two deleted genes that define two connected transcriptome states of the graph are shown at the indicated similarity thresholds. For the classification of yeast genes, four selected categories of each YPD (Costanzo *et al.*, 2000) entry were used. Genes missing from the databases or listed as 'unknown' were excluded from the analyses. ***b***, The list of transcriptomes/deleted genes with the highest number of connections on the similarity graph at the indicated similarity threshold values. Their number of links is given in parenthesis. Those showing the highest connectivity at $C$=0.80 column are colored black, others are listed in gray. Vertical color codes indicate the cellular role categories in the YPD (Costanzo *et al.*, 2000) classification. Metabolism and energy generation (red), DNA/RNA related (yellow), protein synthesis and modification (green), cell stress (magenta), cell cycle, cell fate, mating (blue), signal transduction and transport (gray) and other (light blue) are shown.

**Figure 4** The characterization of links between transcriptome states.

***a***, The descending sequence of transcriptome segment usage frequencies for all the 210 segments ***b***, The average percent ± S.D. of identical transcriptome segments- (red) and identical genes within segments (blue) in any two adjacent links of the transcriptome similarity network, shown at different similarity threshold levels. ***c***, The 10 transcriptome segments used most frequently in establishing links in the transcriptome similarity network, and the 10 genes most frequently dominant in each of them. Vertical color codes indicate the cellular role categories in the YPD (Costanzo *et al.*, 2000) classification, as described in Figure 3.

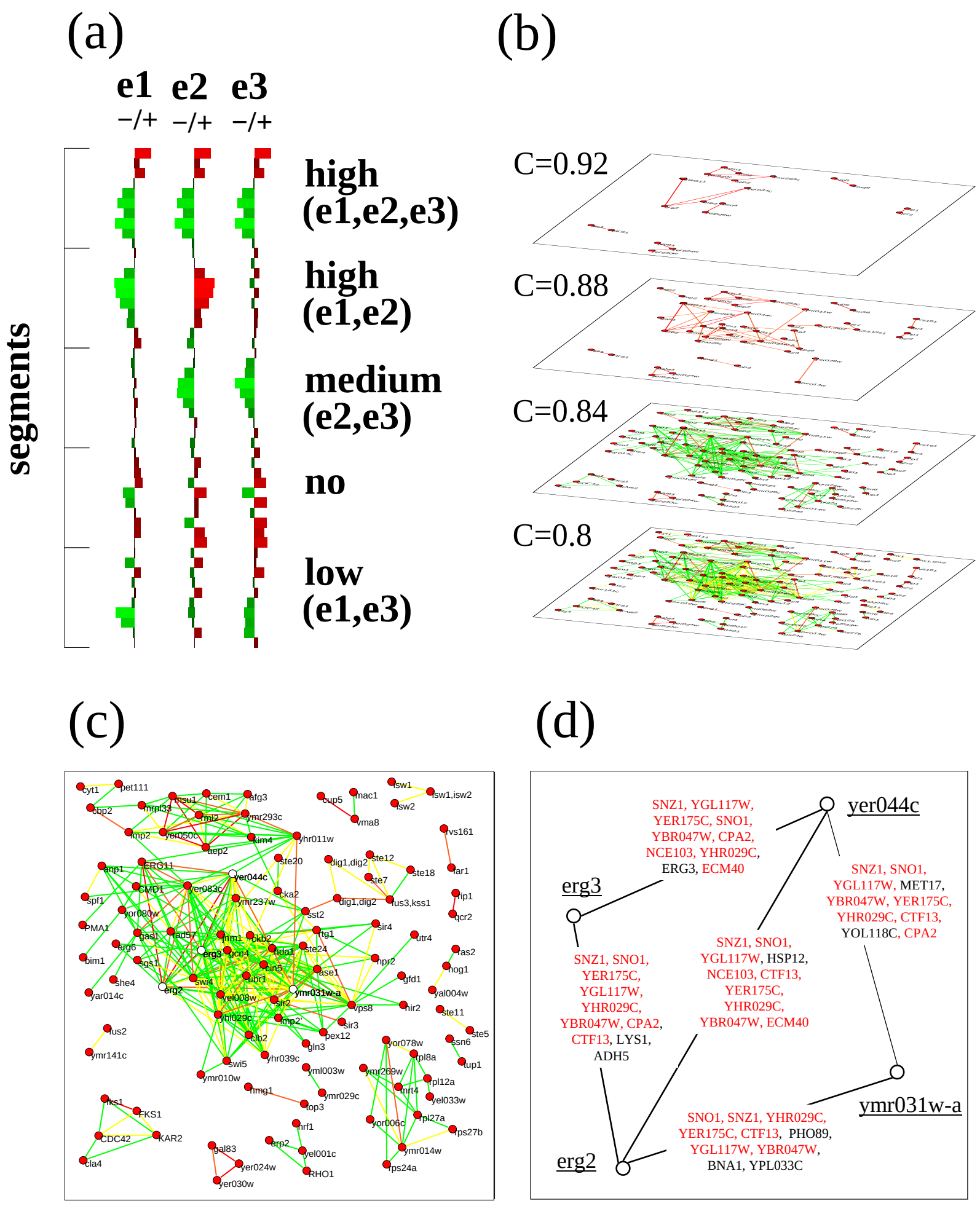
(a)
e1
e2
e3
−/+
−/+
−/+
segments
high
(e1,e2,e3)
high
(e1,e2)
medium
(e2,e3)
no
low
(e1,e3)
(b)
C=0.92
C=0.88
C=0.84
C=0.8
(c)
(d)
SNZ1, YGL117W, YER175C, SNO1, YBR047W, CPA2, NCE103, YHR029C, ERG3, ECM40
yer044c
erg3
SNZ1, SNO1, YGL117W, MET17, YBR047W, YER175C, YHR029C, CTF13, YOL118C, CPA2
SNZ1, SNO1, YGL117W, HSP12, NCE103, CTF13, YER175C, YHR029C, YBR047W, ECM40
SNZ1, SNO1, YER175C, YGL117W, YHR029C, YBR047W, CPA2, CTF13, LYS1, ADH5
ymr031w-a
SNO1, SNZ1, YHR029C, YER175C, CTF13, PHO89, YGL117W, YBR047W, BNA1, YPL033C
erg2

Figure 1.

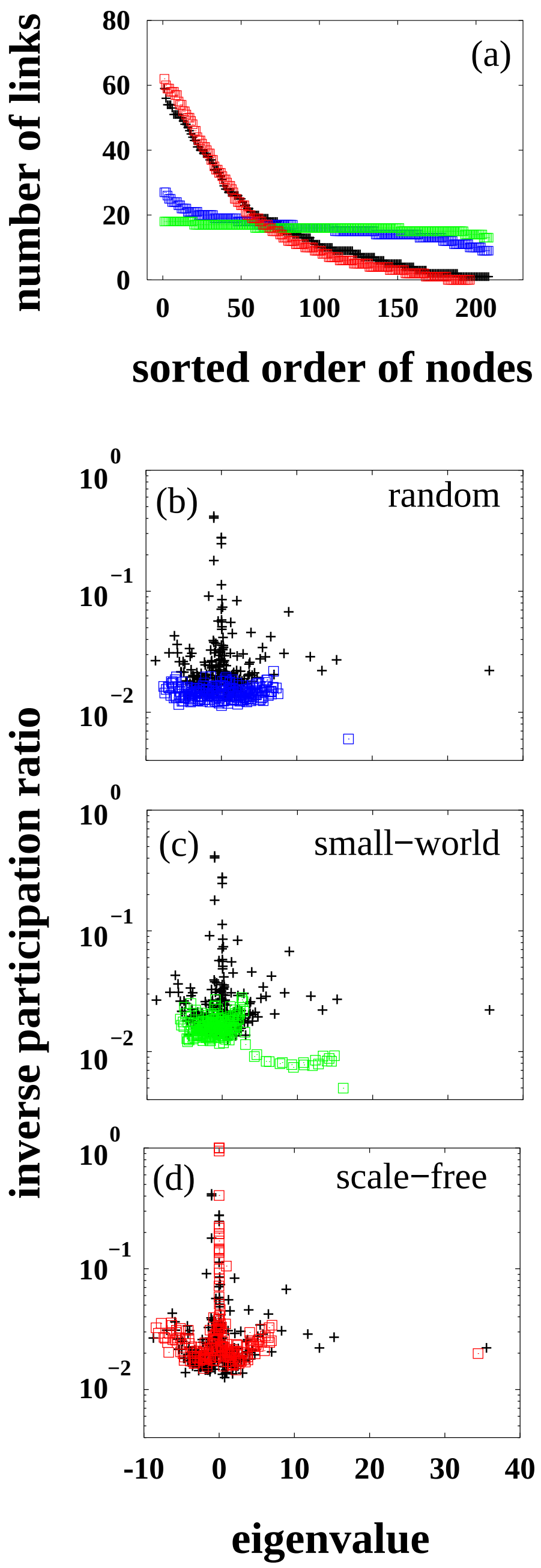

Figure 2.

(a)

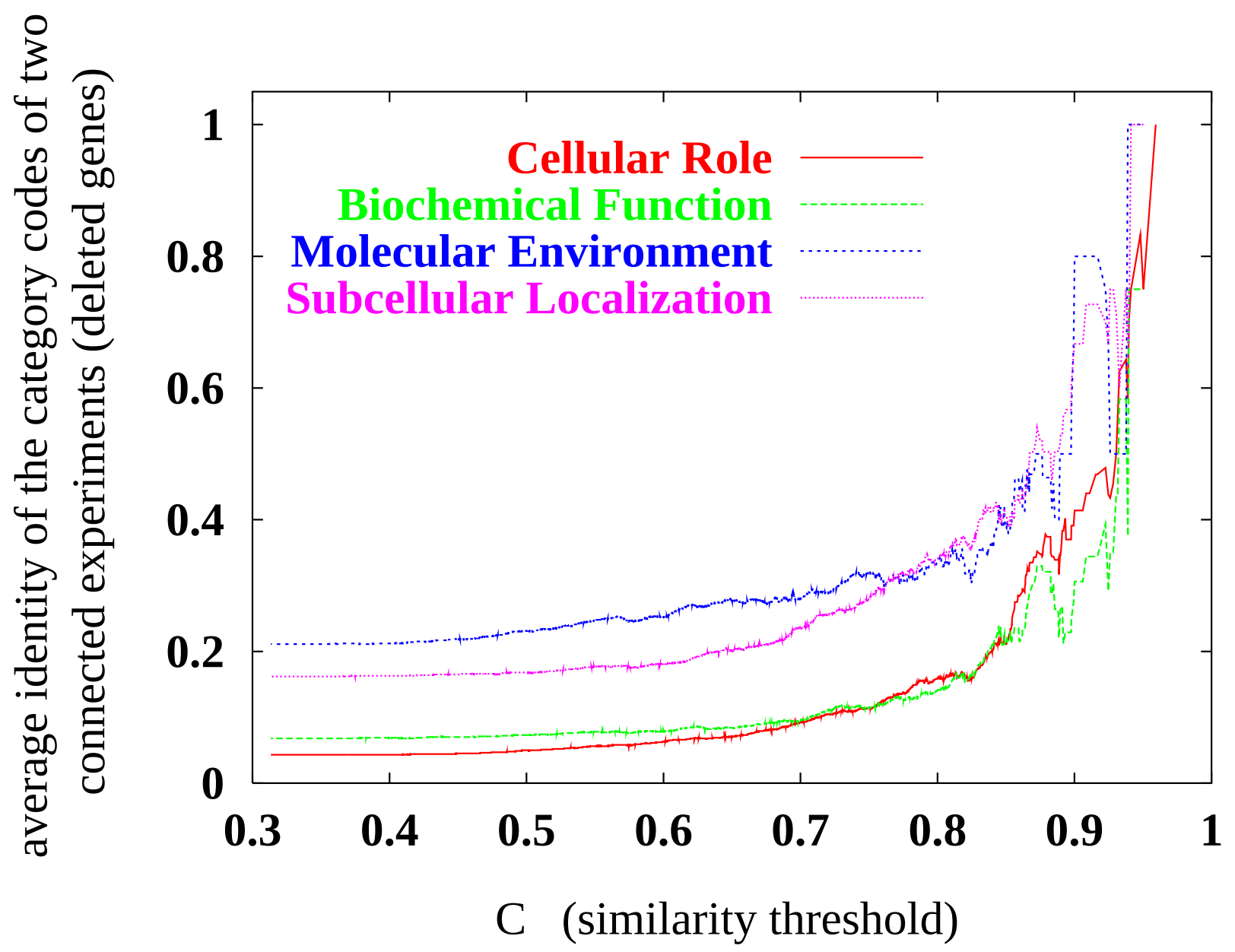

(b)

| C=0.76 | | 0.78 | | 0.80 | | 0.82 | | 0.84 | |
|---|---|---|---|---|---|---|---|---|---|
| yel008w | (37) | sir2 | (29) | swi4 | (24) | swi4 | (19) | ymr031w-a | (16) |
| gcn4 | (35) | swi4 | (28) | ymr031w-a | (23) | ymr031w-a | (18) | gcn4 | (14) |
| yer083c | (35) | gcn4 | (27) | yel008w | (23) | yel008w | (18) | sir2 | (14) |
| swi4 | (34) | ymr031w-a | (27) | yhl029c | (22) | yer083c | (17) | yhl029c | (13) |
| sir2 | (34) | yer083c | (27) | jnm1 | (22) | sir2 | (17) | yel008w | (12) |
| ymr031w-a | (33) | yel008w | (26) | yer083c | (21) | gcn4 | (16) | erg3 | (12) |
| vps8 | (31) | ase1 | (26) | yer044c | (21) | yer044c | (16) | hda1 | (11) |
| ase1 | (31) | yer044c | (25) | sir2 | (21) | yhl029c | (16) | yer044c | (11) |
| yhl029c | (31) | yhl029c | (25) | gcn4 | (20) | erg3 | (15) | swi4 | (11) |
| jnm1 | (31) | jnm1 | (25) | erg3 | (20) | ase1 | (15) | vps8 | (10) |
| hda1 | (30) | hda1 | (24) | ase1 | (19) | hda1 | (15) | ase1 | (9) |
| erg3 | (29) | vps8 | (24) | hda1 | (19) | vps8 | (14) | yer083c | (8) |
| yer044c | (28) | erg3 | (22) | vps8 | (18) | ste24 | (13) | jnm1 | (7) |
| ckb2 | (25) | erg2 | (20) | clb2 | (17) | ubr1 | (12) | erg2 | (7) |
| erg2 | (25) | ubr1 | (20) | ste24 | (17) | jnm1 | (11) | ymr293c | (7) |
| ubr1 | (25) | ste24 | (20) | erg2 | (16) | ymr293c | (9) | ste24 | (5) |
| clb2 | (25) | cin5 | (19) | ubr1 | (15) | erg2 | (9) | ubr1 | (5) |
| ste24 | (24) | clb2 | (19) | rad57 | (14) | yhr011w | (9) | yhr039c | (5) |
| rad57 | (23) | ckb2 | (18) | cin5 | (13) | yhr039c | (7) | yer050c | (5) |
| gas1 | (23) | imp2’ | (17) | yhr011w | (12) | rtg1 | (7) | msu1 | (5) |
| cin5 | (22) | rad57 | (17) | gas1 | (11) | aep2 | (7) | aep2 | (5) |
| imp2’ | (21) | gas1 | (16) | imp2’ | (11) | clb2 | (7) | "dig1,dig2" | (5) |
| ERG11 | (19) | yhr039c | (14) | ERG11 | (10) | msu1 | (7) | rml2 | (4) |
| anp1 | (19) | yhr011w | (14) | yhr039c | (10) | ckb2 | (6) | ERG11 | (4) |
| swi5 | (19) | swi5 | (13) | ymr293c | (10) | rml2 | (6) | clb2 | (4) |

Figure 3.

Figure 4.

(a)

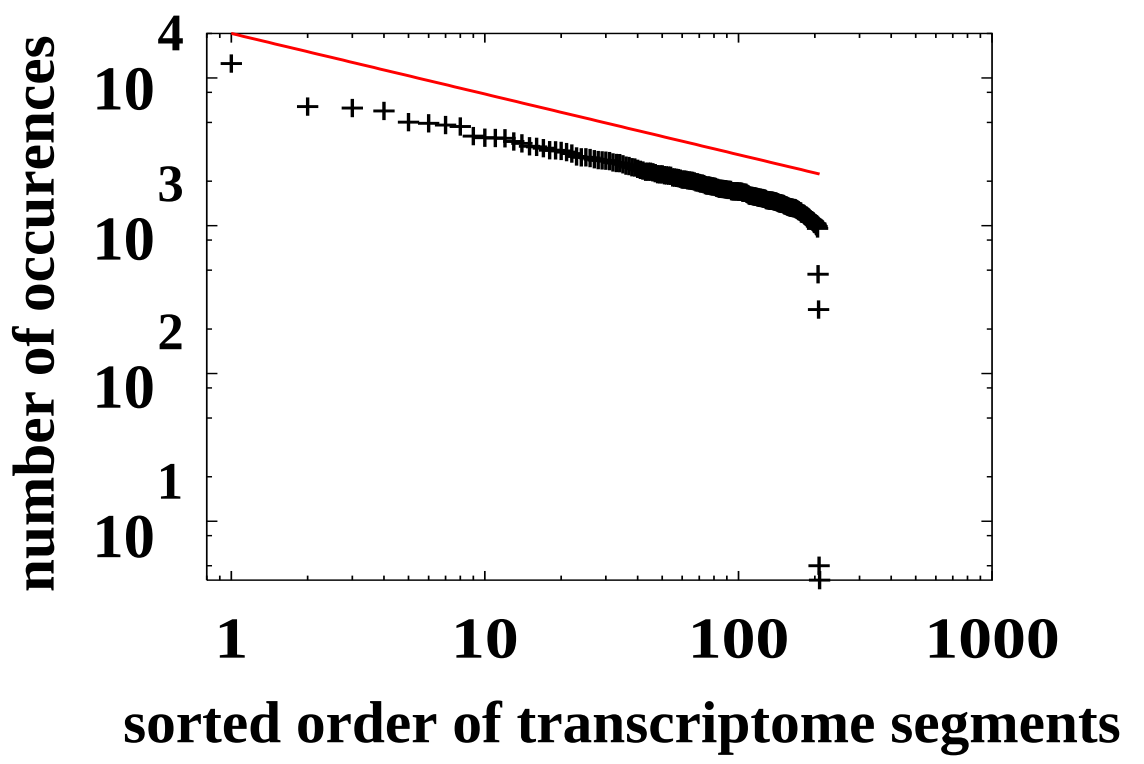

(b)

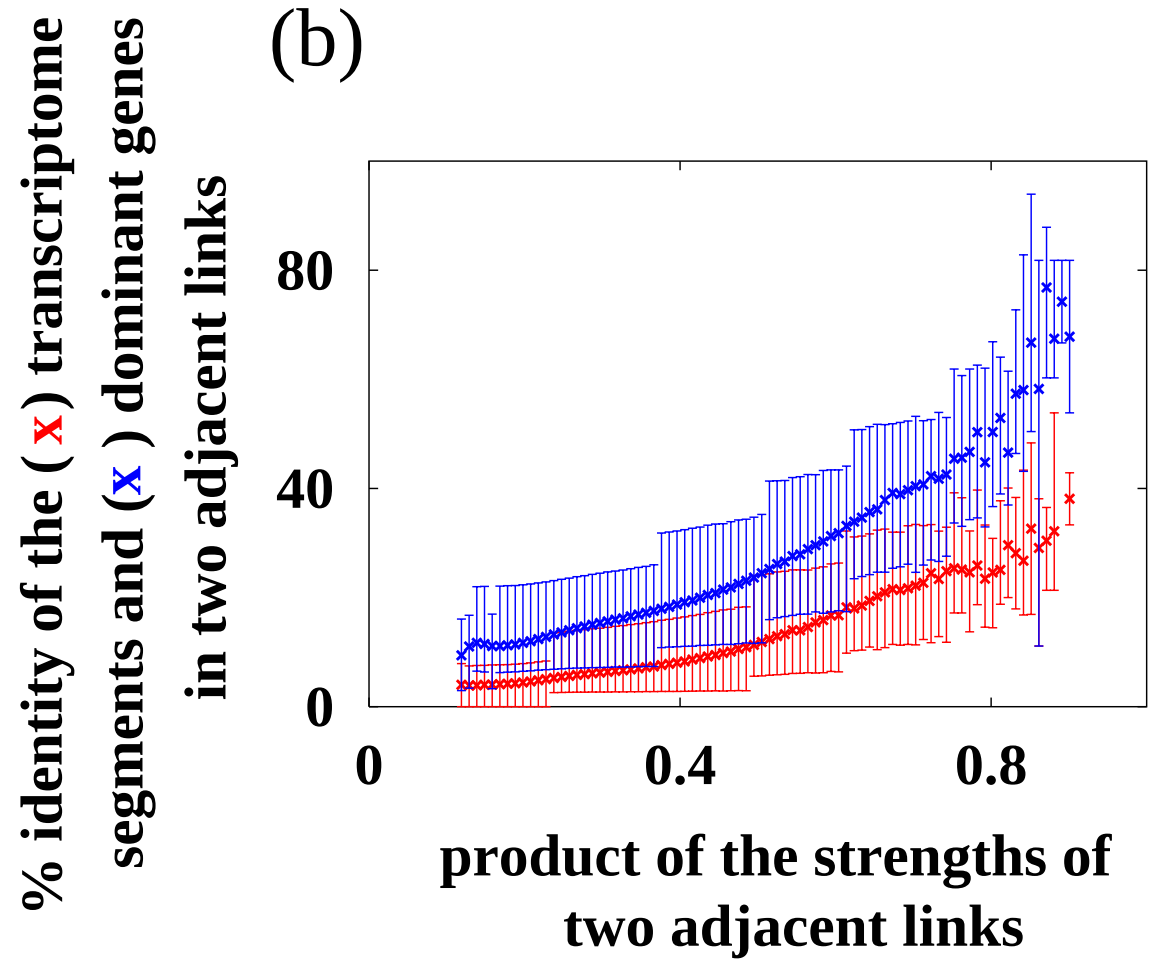

(c)

**Indices of the 10 transcriptome segments occuring most frequently**

**The 10 most strongly coupling genes within each of these transcriptome segments**

| | | | | | | | | | | |
|---|---|---|---|---|---|---|---|---|---|---|
| 145 | spo20 | yml131w | hxt2 | plb1 | erg5 | sok2 | ymr018w | ymr010w | ymr009w | cos3 |
| 59 | yer175c | spi1 | yer160c | spt15 | yer158c | yer152c | yer156c | bem2 | grx4 | pab1 |
| 71 | pox1 | ygl217c | aro8 | mds3 | mig2 | rps26a | clg1 | ygl193c | hos2 | kip3 |
| 142 | yml047c | yml040w | yml045w | cyb2 | cat2 | yml059c | ndc1 | cmp2 | yml037c | rrn11 |
| 87 | yhr079bc | pcl5 | yhr087w | ste12 | yhr085w | yhr088w | ssf1 | rpp1 | gic1 | yhr083w |
| 102 | arg3 | sip4 | pry3 | pry1 | mef2 | yjl086c | bck1 | scp160 | ics3 | yjl083w |
| 13 | ybr147w | adh5 | ics2 | ybr144c | spp381 | ysw1 | ics4 | ara1 | ybr139w | ybr159w |
| 66 | pyc1 | ygl045w | ole1 | ygl039w | ygl046w | ygl059w | pus2 | ygl068w | ygl053w | ygl047w |
| 9 | fig1 | ybr047w | zta1 | gip1 | gal7 | ybr032w | pdx3 | ybr043c | gal10 | chs2 |
| 58 | yer138c | ftr1 | yer124c | yer135c | yer128w | scs2 | yer130c | yer119c | mag1 | ddi1 |